\documentclass[a4paper,11pt,amsmath,amssymb,amsfonts,onecolumn,nofootinbib]{article}
\pdfoutput=1
\usepackage{jheppub}

\usepackage[T1]{fontenc}

\usepackage[normalem]{ulem}

\usepackage{physics}
\usepackage{amsfonts}
\usepackage{amsmath}
\usepackage{placeins}
\usepackage{color}
\definecolor{ao}{rgb}{0, 0.7, 0}
\usepackage{graphicx}
\usepackage{makecell}

\usepackage{hyperref}
\usepackage{array}
\usepackage[shortlabels]{enumitem}

\usepackage{cleveref}
\usepackage{multicol}
\usepackage[nolist,nohyperlinks]{acronym}

\newcommand{\bdy}{\mathrm{bdy}}
\newcommand{\bulk}{\mathrm{bulk}}

\newcommand{\vphi}{\varphi}

\definecolor{orange}{rgb}{0.7,0.35,0}

\begin{document}
\title{Quantization and variational problem of the Gubser-Rocha Einstein-Maxwell-Dilaton model, conformal and non-conformal deformations, and 
its proper thermodynamics.}

\author{Nicolas Chagnet${}^1$, Floris Balm${}^1$, and Koenraad Schalm${}^1$\\
${}^1${\it
Institute Lorentz for Theoretical Physics, $\Delta$-ITP, Leiden University}\\
{\it Niels Bohrweg 2, Leiden, the Netherlands.}\\
E-mail: {\tt chagnet@lorentz.leidenuniv.nl}, {\tt balm@lorentz.leidenuniv.nl}, {\tt kschalm@lorentz.leidenuniv.nl}
}

\abstract{
We show that the strongly coupled field theory holographically dual to the Gubser-Rocha anti-de-Sitter Einstein-Maxwell-Dilaton theory describes not a single non-trivial AdS$_2$ IR fixed point, but a one-parameter family. It is dual to a local quantum critical phase instead of a quantum critical point. This result follows from a detailed analysis of the possible quantizations of the gravitational theory that is consistent with the thermodynamics of the analytical Gubser-Rocha black hole solution. The analytic Gubser-Rocha black hole is only a 2-parameter subset of all possible solutions, and we construct other members numerically. These new numerical solutions correspond to turning on an additional scalar charge. Moreover, each solution has multiple holographic interpretations depending on the quantization chosen. In one particular quantization involving a multitrace deformation the scalar charge is a marginal operator. In other quantizations where the marginal multitrace operator is turned off, the analytic Gubser-Rocha black hole does not describe a finite temperature conformal fluid. 
}

\maketitle

\begin{acronym}
    \acro{rn}[RN]{Reissner-Nordstr\"{o}m}
    \acro{gr}[GR]{Gubser-Rocha}
    \acro{agr}[aGR]{analytical Gubser-Rocha}
    \acro{fg}[FG]{Fefferman-Graham}
    \acro{emd}[EMD]{Einstein-Maxwell-Dilaton}
\end{acronym}

\section{Introduction}

One of the main insights holography has provided into the physics of strongly correlated systems is the existence of previously unknown (large $N$) non-trivial IR fixed points. These fixed points are characterized by an emergent scaling symmetry of the Lifshitz form categorized by a dynamical critical exponent $z$, a hyperscaling exponent $\theta$, and a charge anomalous dimension $\zeta$.
\begin{align}
    x\rightarrow \lambda^{1/z} x~,~t\rightarrow\lambda t~,~F\rightarrow \lambda^{\frac{d-\theta}{z}}F~,~\rho\rightarrow \lambda^{\frac{d - \theta + \zeta}{z}}\rho ~.
\end{align}
Here $F$ is the free energy density and $\rho$ the charge density \cite{Faulkner:2009wj,Charmousis:2010zz,Gouteraux:2011ce,Huijse:2011ef}. Within these Lifshitz fixed points those with $z=\infty$ are special.  Such theories have energy/temperature scaling with no corresponding spatial rescaling. These are therefore systems with exact {\em local quantum criticality}. Phenomenologically this energy/temperature scaling without a corresponding spatial part is observed in high $T_c$ cuprates, heavy fermions and other strange metals, where this nomenclature originates (see e.g. \cite{Si_2001}). In holography $z=\infty$ IR fixed points correspond to an emergent AdS$_2$ symmetry near the horizon of the extremal black hole. The two most well-known such solutions are the plain extremal \ac{rn} black hole and the extremal \ac{gr} black hole \cite{Gubser:2009qt}.
The \ac{rn} solution of AdS-Einstein-Maxwell theory has been studied extensively primarily because it is the simplest such model. Its simplicity also means it is too constrained to be realistic as a model of observed locally quantum critical metals. Notably the \ac{rn} has a non-vanishing ground-state entropy and emerges from a $d>2$-dimensional conformal field theory. The more realistic \ac{gr} model arises from a non-conformal strongly correlated theory, where one isolates the leading irrelevant deformation from the IR fixed point. This ``universal'' subsector gives it a chance to be applicable to observed local quantum critical systems. Moreover the groundstate now has vanishing entropy (to leading order). In the gravitational description this leading (scalar) (IR)-irrelevant operator is encoded in a dilaton field that couples non-minimally to both the Einstein-Hilbert action and the Maxwell action. Even with its more realistic appeal, the more complex nature of the \ac{gr} dynamics means it has been studied less; some examples are 
\cite{Goldstein:2009cv,Ling:2013nxa,Davison:2013txa,Kim:2016dik,Caldarelli:2016nni}.

In the course of these studies of non-minimally coupled \ac{emd} theories, it was noted in particular that the proper holographic interpretation of the \ac{agr} black hole solution depends sensitively on the particular quantization \cite{Kim:2016dik,Caldarelli:2016nni}. Within holography, relevant and marginally relevant scalars allow for different quantization schemes. A relevant operator of dimension $\frac{d}{2} <\Delta < d$ always has a conjugate operator of dimension\footnote{The upper bound of $\Delta$ would suggest $\Delta_{\mathrm{conj}} > 0$ but requiring unitarity of the conjugate theory leads to a higher bound.} $\frac{d}{2} - 1<\Delta_{\text{conj}}=d-\Delta <\frac{d}{2}$, and one can choose whether one considers the original operator as the dynamical variable (standard quantization) or the conjugate operator (alternate quantization) or any intermediate linear combination through a double-trace deformation \cite{Witten:2001ua,Mueck:2002gm}.

An additional complication results from the fact that the (static and isotropic) \ac{agr} solution is a two-parameter solution depending on $T$ and $\mu$, whereas one expects a third independent parameter encoding the asymptotic source value of the dilaton field. A low-energy scalar can have a sourced (or unsourced) vacuum-expectation value; this changes the energy of the ground-state and hence should contribute to the thermodynamics. For minimally coupled scalars this was recently elucidated in \cite{Li:2020spf}.

In this paper we will show that the correct way to interpret the \ac{agr} solution is as a two-parameter subset of solutions within the three-parameter thermodynamic phase diagram.  For essentially all quantization schemes this constrains the source of the dilaton field in terms of the temperature and chemical potential of the solution. Crucially this implies that derivatives of thermodynamic potentials mix the canonical contribution with an additional contribution from the scalar response. We will show this explicitly in Section \ref{sec:choiceofquant}. A proper understanding of the solution requires one to carefully separate out this contribution. 

It also turns out, however, that there is a specific quantization scheme where the dilaton corresponds to an exactly marginal operator in the theory. This was previously noted for another set of the \ac{emd} actions \cite{Caldarelli:2016nni}.\footnote{We thank Blaise Goutéraux for bringing this paper to our attention.} In this special quantization choice the \ac{agr} solution corresponds to a 
solution with no explicit source for the dilaton field. Within this special quantization scheme one can deform the analytical solution to a nearby solution with a finite scalar source. We do so in Section \ref{sec:oneparameterfam}. We conclude with a brief discussion on the meaning of this newly discovered exactly marginal deformation.

\section{Setup}

The \ac{gr} black hole is a solution to
the \ac{emd} action
\begin{equation}
    \label{eq:gravAction}
    S_\bulk = \dfrac{1}{2 \kappa^2} \int \dd^4 x \sqrt{-g} \left[ R - \dfrac{Z(\phi)}{4} F^2 - \dfrac{1}{2} (\partial \phi)^2 - V(\phi) \right]~,
\end{equation}
where the potentials are given by $Z(\phi) = e^{\phi/\sqrt{3}}$ and $V(\phi) = - 6 \cosh(\phi/\sqrt{3})$~.\footnote{Note that the dilaton has dimension zero.} This action is a consistent truncation of $d=11$ supergravity compactified on $AdS_4\times S_7$ \cite{Gubser:2009qt}.  
The equations of motion for this system are
\begin{equation}
    \label{eq:eomGR}
    \begin{aligned}
        R_{\mu\nu} & = \dfrac{Z(\phi)}{2} \left[ F_{\mu}{}^{\rho} F_{\nu \rho} - \dfrac{1}{4} g_{\mu\nu} F^2 \right] + \dfrac{1}{2} \partial_\mu \phi \partial_\nu \phi + \dfrac{1}{2} g_{\mu\nu} V(\phi)~,\\
        \nabla_{\mu} \left[ Z(\phi) F^{\mu\nu} \right] & = 0~,\\
        \Box \phi & = V^\prime(\phi) + \dfrac{Z^\prime(\phi)}{4} F^2~,
    \end{aligned}
\end{equation}
where we used that, on-shell, $R = 2 V(\phi) + \dfrac{1}{2} (\partial \phi)^2$. The static and isotropic metric ansatz that is asymptotically AdS is
\begin{equation}
\label{eq:metric}
    \dd s^2 = g_{\mu\nu} \dd x^\mu \dd x^\mu = \dfrac{1}{z^2} \left[ - f(z) \dd t^2 + g(z) \left( \dd x^2 + \dd y^2 \right) + \dfrac{\dd z^2}{f(z)} \right]~, 
\end{equation}
where the coordinate $z$ is the radial direction with $z = 0$ the AdS boundary (UV). 
The \ac{agr} solution \cite{Gubser:2009qt} is then given by
\begin{equation}
    \label{eq:homSolution}
    \begin{aligned}
        g(z) & = (1 + Q z)^{3/2}~,\\
        f(z) & = \dfrac{1- z/z_h}{g(z)} \left[ 1 + \left(1 + 3 Q z_h\right)  \dfrac{z}{z_h} + \left(1 + 3 Q z_h + 3 Q^2 z_h^2\right) \left(\dfrac{z}{z_h}\right)^2 \right]~,\\
        A_t(z) & = \mu j(z) =  \frac{\sqrt{3 Q z_h (1 + Q z_h)}}{z_h} \dfrac{1 - z/z_h}{1 + Q z}~, \quad \, 
        \\
        \phi(z) & = \dfrac{\sqrt{3}}{2} \log\left[ 1 + Q z \right]~,
    \end{aligned}
\end{equation}
where $z_h$ is the horizon  of this non-extremal black hole. From hereon we choose units where $2\kappa^2=16\pi G=1$, such that the temperature, chemical potential and entropy-density of the \ac{gr}-black hole are
\begin{equation}
    \begin{gathered}
    \label{eq:GCmatching}
    T = \left. -\frac{f'(z)}{4\pi}\right|_{z=z_h} = \dfrac{3 \sqrt{1 + Q z_h}}{4 \pi z_h} ~,~~ s= 4 \pi a_h= 4 \pi\dfrac{\left( 1 + Q z_h \right)^{3/2}}{z_h^2}~,\\~~ \mu = A_t(z=0) = \sqrt{3 Q z_h (1 + Q z_h)}/z_h~,
\end{gathered}
\end{equation}
where $a_h = \sqrt{g_{xx}(z_h) g_{yy}(z_h)}$ is the area density of the horizon.
Expressed in terms of the temperature, it is easy to see that the entropy vanishes linearly $s =\frac{16\pi^2}{3\sqrt{3}}\mu T+\ldots$ at low temperatures with no remnant ground state entropy.
Important in the remainder is (1) to recall that both the temperature and the entropy can be read off from the near-horizon behavior of the metric alone. As local properties of the black hole they do not depend on the boundary conditions. (2) The analytic solution depends on two parameters $Q$ and $z_h$. 
And (3) note that the metric gauge choice is not of the \ac{fg} type in that the change in metric functions starts at order $z$ and not $z^3$.

\section{Regularization, boundary terms and choice of quantization}

\subsection{Boundary action}
\label{sec:boundary-action}
We must add to the gravitational action \eqref{eq:gravAction} a boundary action. This is to regularize its on-shell value as well as  to make the variational principle well-defined. In the case of the scalar it also  prescribes the quantization of the scalar field. 
We will be using in this work a standard multi-trace deformation of the Neumann boundary theory, which were generally described in \cite{Witten:2001ua,Mueck:2002gm,Papadimitriou:2007sj} and more specifically in \ac{emd} theories \cite{Caldarelli:2016nni}, with a boundary action of the form
\begin{align}
    \label{eq:bdyAction}
    S_\bdy = -\int_{z = \epsilon} \dd^3 x \sqrt{-\gamma} \left[ 2 K + 4 + {}^{(3)}\! R_\gamma\right] + S_{\bdy,\phi}~,
\end{align}
Here $N^\mu = -\sqrt{g^{zz}} (0, 0, 0, 1)$ is an outward pointing spacelike unit normal vector defining the hypersurface $z = \epsilon \ll z_h$ and $\gamma_{\mu\nu} = g_{\mu\nu} - N_\mu N_\nu$ is the induced metric on the surface. Furthermore $K \equiv \gamma^{ij} K_{ij}$ is the trace of the extrinsic curvature $K_{ij} \equiv -\gamma_i^\mu \gamma_j^\nu \nabla_{(\mu} N_{\nu)}$ and  ${}^{(3)}\!R_\gamma$ the Ricci scalar curvature of the hypersurface (Latin symbols correspond to coordinates on the hypersurface while the greek symbols are those of the original manifold). The first three terms correspond to the usual Gibbons-Hawking-York counterterms necessary to make the variational principle for the metric well-defined and also to regularize the Einstein-Hilbert-Cosmological Constant part of the action on shell. In our coordinatization Eq.~\eqref{eq:metric} the induced metric is flat on-shell. The scalar part of the boundary term $S_{\bdy,\phi}$ can take two forms depending on whether we consider the standard quantization boundary theory where only the $\phi^2$ regularization term appears
\begin{equation}
    \label{eq:bdyPhiSQ}
    S^{(\mathrm{SQ})}_{\bdy,\phi} = \int_{z = \epsilon} \dd^3 x \sqrt{-\gamma} \dfrac{\Lambda_\phi}{2} \phi^2~, \quad \Lambda_\phi = -1~,
\end{equation}
--- here the value of $\Lambda_\phi$ is set to regularize the boundary term arising from varying the bulk action --- or whether we consider a multi-trace deformation of the alternate quantization boundary theory
\begin{equation}
    \label{eq:bdyPhiMT}
    S^{(\mathrm{MT})}_{\bdy,\phi} = \int_{z = \epsilon} \dd^3 x \sqrt{-\gamma} \left[ \dfrac{\Lambda_\phi}{2} \phi^2 + \phi N^\mu \partial_\mu \phi \right]+ S_F~, \quad \Lambda_\phi = 1~.
\end{equation}
The $\phi N^\mu \partial_\mu \phi$ is a Legendre transform from Dirichlet to Neumann boundary conditions, which also diverges at leading order and is the reason for the shift in $\Lambda_\phi$ as we will see.\footnote{Strictly speaking $\phi N^\mu \partial_\mu \phi$ is a combination of a true Legendre transform $J{\cal O}= z^{\lambda_--\lambda_+-1}\phi\partial_n z^{-\lambda_-}\phi$ (see Eq.~\eqref{eq:scalarExp}) and counterterms.}
The multi-trace deformation $S_F$ is a finite contribution to the boundary action and will be described when the asymptotics of the solution are analysed. We will 
continue the derivation with the choice $S_{\bdy,\phi} = S_{\bdy,\phi}^{(\mathrm{MT})}$ while keeping in mind that a similar derivation can easily be done using instead $S_{\bdy,\phi} = S_{\bdy,\phi}^{(\mathrm{SQ})}$, and we will invoke those results when necessary.

Varying the total action $S = S_\bulk + S_\bdy$ to first order, a proper holographic interpretation demands that one obtains a variation of the form \cite{Balasubramanian:1999re}
\begin{equation}
    \label{eq:actionVariation}
    \delta S = \int_{z = \epsilon} \dd^3 x \sqrt{-\gamma} \left[ \dfrac{1}{2} T_{\mu \nu}\delta \gamma^{\mu \nu} +  J^\mu \delta A_\mu + \mathcal O_{\vphi} \delta \vphi  \right]~,
\end{equation}
where the terms multiplying the \ac{emd} fields are interpreted as the operators in the boundary CFT where $T_{\mu\nu}$ is the boundary stress tensor, $J_\mu$ the boundary current associated with the U(1) charge, and $\mathcal O_\vphi$ the operator dual to a scalar which may be a non-linear function of the dilaton field. The important point is that the action evaluated on the black hole solution is equated with (minus) its Gibbs free energy density. The variation of the action (restricted to preserve isotropy) thus includes  thermodynamic variations. The expression above makes clear that in addition to the temperature and the chemical potential there ought to be a dependence of the Gibbs free energy on an external (source) variation of (the boundary value of) the scalar field \cite{Li:2020spf}.

Performing this variation on Eqs \eqref{eq:gravAction} plus \eqref{eq:bdyAction}, we can write it as a bulk integral of an integrand proportional to the equations of motion \eqref{eq:eomGR}, that vanishes on-shell, and a remaining boundary part.
In the boundary part the normal derivatives of $\delta \gamma_{\mu\nu}$ 
cancel due to the Gibbons-Hawking-York term; there are no normal derivatives in $A_\mu$.
Restricting to boundary indices we have\footnote{The radial components of $T_{\mu\nu}$ and $J_\mu$ vanish due to the projection on the hypersurface.}
\begin{equation}
    \label{eq:vevs}
    \begin{aligned}
        T_{ij} & = 2K_{ij} - 2\,({}^d R_{\gamma,ij}) - 2(K + 2) \gamma_{ij} +\gamma_{ij} \left[  \phi N^z \partial_z \phi + \Lambda_\phi \phi^2/2 \right] + T_{ij}^F~,\\
        J_i & = - Z(\phi) N^z F_{z i}~,
    \end{aligned}
\end{equation}
where $T_{ij}^F$ is the contribution from $S_F$. The expression for $\mathcal O_\vphi$ requires a more detailed discussion. Focusing on the variation in the dilaton $\phi$ in \eqref{eq:actionVariation}, we have
\begin{equation}
    \label{eq:bdyActionScalar}
    \delta S_\phi = \int_{z = \epsilon} \dd^3 x \sqrt{-\gamma} \left[ \Lambda_\phi \phi \delta \phi   + \phi N^z \partial_z \delta \phi  \right] + \delta S_F~.
\end{equation}
From its linearized equation of motion the dilaton has the following expansion in the near-boundary region
\begin{equation}
    \label{eq:scalarExp}
    \phi(z) =\alpha z^{\lambda_-} + \beta z^{\lambda_+} + \mathcal O(z^3)~,
\end{equation}
where $\lambda_\pm = \dfrac{3}{2} \pm \dfrac{1}{2} \sqrt{9 + 4 m^2}$ and $m$ is the effective mass. In the \ac{gr} model the effective mass equals
\begin{equation}
    m^2 = \left. \dfrac{\partial}{\partial \phi^2} \left[ V(\phi) + \dfrac{Z(\phi)}{4} F^2 \right] \right|_{\phi = 0, z \to 0} = -2~.
\end{equation} 
This value of the mass $-\frac{9}{4}<m^2<1-\frac{9}{4}=-\frac{5}{4}$ is in the regime where two different quantizations are allowed, i.e. for this value of $m$ both $\lambda_\pm >0$ and either $\alpha$ (standard) or $\beta$ (alternate) can be chosen as the source for the dual CFT operator with the other the response. 
One can also choose a mixture of the two, corresponding to a multi trace deformation, as we shall elucidate below.

The proper holographic normalization is most conveniently performed in a \ac{fg} ansatz for the metric
\begin{equation}
    \label{eq:fg-ansatz}
    \dd s^2 = \dfrac{1}{z^2} \left[ -H_{tt}(z) \dd t^2 + H_{xx}(z) \dd x^2 + H_{yy}(z) \dd y^2 +  \dd z^2 \right]~,
\end{equation}
where we require Anti-deSitter (AdS) aymptotics $H_{\mu\nu}(z = 0) = 1$ and use the equations of motion \eqref{eq:eomGR} to constrain the near-boundary expansion of $H_{\mu\nu}$ in terms of a small subset of degrees of freedom. We will use this ansatz for the remainder of this section. Using that $N^z(z) = -z$, and substituting \eqref{eq:scalarExp} into \eqref{eq:bdyActionScalar}, we can expand the variation w.r.t. the dilaton as
\begin{equation}
        \delta S_\phi = \int_{z = \epsilon} \dd^3 x \left[ \dfrac{\Lambda_\phi - 1}{\epsilon} \alpha \delta \alpha + \alpha \delta \beta (\Lambda_\phi - 2) + \beta \delta \alpha (\Lambda_\phi - 1)  + \mathcal O(\epsilon) \right] + \delta S_F~.
\end{equation}
As we claimed in \eqref{eq:bdyPhiMT}, we must remove the leading divergence by imposing $\Lambda_\phi = 1$, leaving a finite contribution
\begin{equation}
    \label{eq:bdyActionScalarExp}
    \delta S_\phi = \int_{z = \epsilon} \dd^3 x \left[ - \alpha \delta \beta  + \mathcal O(\epsilon) \right] + \delta S_F~.
\end{equation}
For the standard quantization term \eqref{eq:bdyPhiSQ}, it is easy to see that a similar derivation leads to $\Lambda_\phi = -1$. 

One can modify the quantization by the addition of a multitrace deformation. This can in general be encoded in the
boundary action $S_F$. Following \cite{Papadimitriou:2007sj,Vecchi:2010dd,Caldarelli:2016nni}, we choose $S_F = \int \dd^3 x \sqrt{-\gamma} \epsilon^d \mathcal F(\alpha)$ such that, ignoring the metric variation, $\delta S_F = \int \dd^3 x \sqrt{-\gamma} \epsilon^d \mathcal F^\prime(\alpha) \delta \alpha$.
Without loss of generality we choose $\mathcal F$ of the form 
$\mathcal F(\alpha) = \frac{a}{2} \alpha^2 + \frac{b}{3} \alpha^3$ from here on.
The variation of the boundary action then becomes
\begin{equation}
    \delta S_\phi = \int_{z = \epsilon} \dd^3 x \, \alpha \left[ - \delta \beta + (a + b \alpha) \delta \alpha\right]~.
\end{equation}
We can therefore identify the VEV of the boundary scalar operator as $\mathcal O_\varphi = \alpha$ while the source of the operator is 
\begin{equation}
    \label{eq:boundary-condition}
    J_{\mathrm{MT}} = -\beta + a \alpha + \frac{b}{2} \alpha^2~.
\end{equation}
Once again, had we chosen the standard quantization boundary term, then we would have $\delta S_\phi = \int \dd^3 x \beta \delta \alpha$ such that $\mathcal O_\varphi = \beta$ and $\varphi = \alpha$ leading to the boundary condition $J_{\mathrm{SQ}} = \alpha$.

We have now almost all the ingredients to compute the scalar contribution to the stress tensor, but we still need to derive the variation of $S_F$ w.r.t. the leading order of the boundary metric in order to compute the term $T_{ij}^F$, as was done before in \cite{Caldarelli:2016nni}. Doing so, one simply finds $T^F_{ij} = \gamma_{ij} \epsilon^d \mathcal F(\alpha)$. It is interesting to note that the contribution $S_F$ can also be absorbed into corrections to the $\phi^2$ term as well as a $\phi^3$ term as
\begin{equation}
    S_{\bdy} = \int_{z = \epsilon} \dd^3 x \sqrt{-\gamma} \left[ -(2 K + 4 + {}^{(3)}\! R_\gamma) + \dfrac{\Lambda_\phi + \epsilon a}{2}  \phi^2 + \phi N^\mu \partial_\mu \phi + \frac{b}{3} \phi^3 \right]~,
\end{equation}
where $\Lambda_\phi + \epsilon a$ is a renormalized $\phi^2$ coupling which will reproduce the $\alpha^2$ contribution of $\mathcal F$, as was done in e.g. \cite{Witten:2001ua,Bernamonti:2009dp}. 
The $\phi^3$ coupling on the other end will reproduce the $\alpha^3$ contribution of $\mathcal F$. 
This way of writing the boundary action action highlights why 
we concentrated on $\mathcal F$ of the form $\mathcal{F}(\alpha)=\frac{a}{2}\alpha^2+\frac{b}{3}\alpha^3$.
Lower order in $\alpha$ terms are constant shifts variationally and can be absorbed in a field redefinition -- they are tadpoles.
Any term $\alpha^n$
 for $n > d$ would lead to vanishing contributions $\epsilon^{n-d}$ in the action -- they are irrelevant deformations.
The equality $\Lambda_\phi = 1$ remains true in order to regularize $\delta S$.

In the presence of such a boundary action, the contribution $T_{ij}^F$ in the expression \eqref{eq:vevs} 
simply includes the $\phi^2,\phi^3$ contributions and leads to
\begin{equation}
    T_{ij} = 2K_{ij} - 2\,({}^d R_{\gamma,ij}) - 2(K + 2) \gamma_{ij} +\gamma_{ij} \left[  \phi N^z \partial_z \phi + \frac{\Lambda_\phi + \epsilon a}{2} \phi^2 + \frac{b}{3} \phi^3 \right]~.\\
\end{equation}
We recognize the $\mathcal F$-dependent part of the stress tensor which agrees with the direct method. It is then immediate to compute the trace of the stress tensor
\begin{equation}
    \label{eq:traceSTgeneral}
    T_i{}^i = \frac{\alpha}{2} \left( 3 a \alpha + 2 b \alpha^2 -4 \beta \right) = -\frac{\alpha}{2} \left( a \alpha - 4 J_{\mathrm{MT}} \right)~,
\end{equation}
where in the last equality we used the boundary condition \eqref{eq:boundary-condition}. 
This result points to the existence of a line of critical points with $a = 0$ where the sourceless ($J_{\mathrm{MT}} = 0 $ equivalent to the boundary condition $-\beta+a\alpha+\frac{b}{2}\alpha^2=0$) deformation $\mathcal F$ is just marginal. This is equivalent to only deforming the boundary theory through a 
$\phi^3$ term which indeed has dimension $d$ and should therefore be marginal.

For completeness we mention that in the case of the standard quantization 
the trace of the stress tensor is simply $T_i{}^i = \alpha \beta 
= \beta J_{\mathrm{SQ}}$.

\subsection{Choice of quantization and thermodynamics}
\label{sec:choiceofquant}

In this subsection, we will derive the thermodynamics of a black hole solution in a general compatible quantization choice. This goes beyond the analyses in \cite{Kim:2016dik,Caldarelli:2016nni} where only the thermodynamics of a marginal scalar were considered, i.e. the case of alternate quantization with a multitrace deformation such that the stress tensor remains traceless. In view of extending the choice of possible theories to non-marginal ones, we will show that the thermodynamics space is extended from a 2-parameter to a 3-parameter space, as also emphasized for Einstein-Scalar theory in \cite{Li:2020spf}.

Let us start with the constraint that a choice of solution imposes on the possible quantization schemes. Indeed, while the choice of boundary terms in the action and therefore of the boundary deformation is a priori agnostic of a given solution to the bulk equations of motion, we have seen that the multi-trace deformation leads to a specific choice of boundary condition on the scalar \eqref{eq:boundary-condition}. Not every solution to the bulk equations of motion \eqref{eq:eomGR} are compatible with every possible boundary condition, as was noted in \cite{Caldarelli:2016nni,Ren:2019lgw}. In the case of the metric corresponding to the \ac{agr} solution \eqref{eq:homSolution}, the scalar $\phi$ has the following falloffs
\begin{equation}
    \label{eq:scalarToFG}
    \phi \sim \alpha z + (\beta - f'(0) \alpha/2) z^2 = \alpha z + (\beta - 3 Q \alpha/4) z^2~,
\end{equation}
where we have related the values of $\phi^\prime(0), \phi^{\prime \prime}(0)$ to the falloffs $\alpha,\beta$ in the \ac{fg} ansatz \eqref{eq:fg-ansatz}. This matching is made explicit in Section~\ref{sec:match-fg}. 
Comparing with the full solution \eqref{eq:homSolution}, we can therefore equate $\alpha = \sqrt{3}Q/2$ and $\beta = \sqrt{3} Q^2/8$. 
Consider then alternate quantization deformed by an arbitrary (relevant and marginal) multitrace deformation.
In that case the source equals
\begin{equation}
    \label{eq:compatibility-general}
    J_{\mathrm{MT}}(Q) = \dfrac{\sqrt{3} Q}{8} \left( 4 a + (\sqrt{3} b - 1)Q \right)~.
\end{equation}
From this equation, we see there are a few distinct cases to consider
\begin{enumerate}[(i)]
    \item \label{list:marginal} $a = 0, b = b_{\mathrm{aGR}} \equiv 1/\sqrt{3}$: every instance of the 2-parameter \ac{agr} solution \eqref{eq:homSolution} is compatible with this choice and is sourceless $J = 0$. This is the sourceless marginal deformation we previously mentioned and which was studied in \cite{Kim:2016dik,Caldarelli:2016nni,Ren:2019lgw}. From Eq.~\eqref{eq:traceSTgeneral}, we see that this boundary theory has $T_i{}^i = 0$.
    
    \item \label{list:standard} $a = 0, b = 0$: the quantization procedure is conventional alternate quantization. In this case, since the solution \eqref{eq:homSolution} is not sourceless, we must impose a Neumann boundary condition $\beta = -J$ with fine-tuned source $J(Q) = -\sqrt{3}Q^2/8$. The explicit source leads to an explicitly broken conformal symmetry in the boundary. (A similar argument holds for standard quantization with a Dirichlet boundary condition $\alpha = J$. One would then need to consider the boundary term $S_{\bdy,\phi} = S_{\bdy,\phi}^{(\mathrm{SQ})}$ instead, 
    and a fine-tuned source $J(Q) = \sqrt{3}Q/2$. Also here the explicit source leads to an explicitly broken conformal symmetry in the boundary.)
    
    \item \label{list:generic} For all the other cases, one can look for explicitly sourced solutions $J = J(Q, a, b)$ defined in Eq.~\eqref{eq:compatibility-general}.\footnote{If we insist on looking for solutions with $J = 0$, one of the couplings $a$ or $b$ must be fine-tuned e.g., $b(Q) = \frac{1}{\sqrt{3}}(1- 4 a /Q)$. As it was noted in \cite{Ren:2019lgw}, this means that fixing $a,b$ to some constant will restrict the space of solutions to those for which $Q = \frac{4 a}{1 - \sqrt{3} b}$. Allowing for a finite, albeit fine-tuned, source $J = J(Q)$ leads to the same result and we will choose this more natural point of view.} This case is fundamentally similar to the case \ref{list:standard}, with the explicit sourcing leading to a non-zero trace of the boundary stress-tensor.
\end{enumerate}
In the end, we see that the only natural sourceless description we have of the solutions \eqref{eq:homSolution} corresponds to the marginal multi-trace deformation, case \ref{list:marginal}. The other cases, \ref{list:standard} and \ref{list:generic}, are better understood as explicitly sourced deformations where the source is fine-tuned to select a certain subset of solutions at a fixed $Q$.

An important aspect is that even though a bulk solution may have different interpretations depending on the quantization choices set out above, the thermodynamics does know about the quantization choice.
Let us 
consider
the free energy of the solutions \eqref{eq:homSolution}. Substituting the solution into the action, the free energy density $\Omega$ of the \ac{agr} black hole solution with compatible boundary condition is given by
\begin{equation}
    \label{eq:GRFE}
    S^{\mathrm{regularized}}_{\mathrm{on-shell}} = - \int \dd^3 x \, \Omega ~, \quad \text{ so } \quad \Omega = -\left(\dfrac{1}{z_h} + Q\right)^3 + \frac{Q^2}{8} \left( Q(1 - \sqrt{3}b) - 3a \right)~.
\end{equation}
Furthermore, the holographic dictionary tells us that the chemical potential and the temperature of the boundary theory are given by \eqref{eq:GCmatching}.
One might be inclined to use this to deduce a variation of $\Omega$ in the 2-parameter grand canonical ensemble $\dd \Omega = - s_1 \dd T - \rho_1 \dd \mu$ and derive from it the thermodynamic entropy and charge density of the theory
\begin{equation}
    \label{eq:2thermo}
    s_1 = - \left( \dfrac{\partial \Omega}{\partial T} \right)_\mu~, \quad \rho_1 = - \left( \dfrac{\partial \Omega}{\partial \mu} \right)_T~.
\end{equation}
However, we have seen from Eq. \eqref{eq:actionVariation} that the free energy variation in the presence of an explicit source should be corrected by a scalar contribution of the form (see also \cite{Li:2020spf})
\begin{equation}
    \label{eq:firstLaw}
    \dd \Omega = -s_2 \dd T - \rho_2 \dd \mu - \mathcal O_\varphi \dd J~.
\end{equation}
This is the full 3-parameter thermodynamics of the system. The fact that the free energy \eqref{eq:GRFE} of the \ac{agr} solution only depends on $T$ and $\mu$, and not on the value of the scalar source means that the \ac{agr} solution should be seen as a 2-parameter constrained solution within this 3-parameter space. This family of solutions is only a subset of all the possible ones for {\em any} given compatible quantization scheme. A direct corollary is that to explore only this analytical set of solutions, variations of $J, T, \mu$ are not independent. Denoting $J$ as the dependent variable, i.e. it is not independent but is a function of both $T$ and $\mu$, then the grand canonical potential varies as
\begin{align}
    \dd \Omega = -\left(s_2+\mathcal O_\varphi \frac{\partial J(T,\mu)}{\partial T}\right) \dd T - \left(\rho_2+\mathcal O_\varphi \frac{\partial J(T,\mu)}{\partial \mu} \right)\dd \mu 
\end{align}
if one constrains one's considerations to \ac{agr} solutions only.

The precise relation of the VEV $\mathcal O_\varphi$ and the source $J$ to the fall-off of the dilaton depends on the quantization scheme as we have just reviewed. A choice of quantization is not a canonical transformation, as shown by \cite{Li:2020spf} in the standard quantization case for Einstein-Scalar theories. Therefore the value of the free energy will depend on this choice. 
This is evident in the dependence on $a,b$ in Eq.~\eqref{eq:GRFE}.
In the full 3-parameter space of solutions this quantization choice dependence would only appear in the dilaton contribution part. In the constrained 2-parameter space of solutions, it would appear to imply that now also the thermodynamic entropy $s_1$ and charge density $\rho_1$ deduced from Eq.~\eqref{eq:2thermo} depend on the quantization, as 
\begin{equation}
\label{eq:NaiveEntropyChargeDensity}
\begin{aligned}
    s_1 & = 4 \pi \dfrac{(1+ Q z_h)^{3/2}}{z_h^2} \left[ 1 + \dfrac{Q^2 z_h^3}{8 (1+Q z_h)^3} \left( Q(1 - \sqrt{3}b) - 2 a \right) \right]~,\\
    \rho_1 & = \mu \dfrac{1 + Q z_h}{z_h} \left[ 1 - \dfrac{Q z_h^2 (2 + Q z_h)}{8 (1+Q z_h)^3}  \left( Q(1 - \sqrt{3}b) - 2 a \right) \right]~.
\end{aligned}
\end{equation}
This is strange, as the Bekenstein-Hawking entropy and the charge density -- the VEV of the sourced gauged field -- are properties of the black hole solution and do not depend on the boundary action which sets the quantization. Indeed they can be read off directly from the geometry as
\begin{equation}
    \label{eq:BlackHoleEntropyChargeDensity}
    \begin{aligned}
        & s_2 = 4 \pi \sqrt{g_{xx}(z_h) g_{yy}(z_h)} = \dfrac{4 \pi (1+ Q z_h)^{3/2}}{z_h^2} & \text{ the area of the horizon of the black hole,}\\
        & \rho_2 = - \partial_z A_t(z \to 0) = \mu \frac{(1 + Q z_h)}{z_h} & \text{ the global U(1) charge.}
    \end{aligned} 
\end{equation}
The solution is of course that in the constrained system $s_1$ and $\rho_1$ are not the true entropy and charge density, as they include the artificial contribution from varying $J(T,\mu)$ following from the constraint to stay within the 2-parameter \ac{agr} solution space. It is then a rather straightforward computation to connect Eqs.~\eqref{eq:NaiveEntropyChargeDensity} and \eqref{eq:BlackHoleEntropyChargeDensity} through the variation of $J$ expressed in Eq.~\eqref{eq:firstLaw}. To that end, we can remember that the source $J$ is constrained by the boundary condition \eqref{eq:compatibility-general} and that in our choice of quantization, we always have $\mathcal O_\varphi = \alpha$. 
In summary, the geometric expressions for the entropy and charge of the \ac{agr} solution are always the correct ones. The difference from the quantities computed from the Gibbs potential can be attributed to the fact that one considers a constrained system: the expression $s_1 = -\left( \frac{\partial \Omega}{\partial T} \right)_{\mu} = -\left( \frac{\partial \Omega}{\partial T} \right)_{\mu} - \mathcal O_\varphi \left( \frac{\partial J}{\partial T} \right)_{\mu}$ contains a term that is absent in the correct definition of the entropy $s_2=-\left( \frac{\partial \Omega}{\partial T} \right)_{\mu,J}$, and similarly for $\rho$.

\begin{figure}
     \centering
     \includegraphics[scale=0.75]{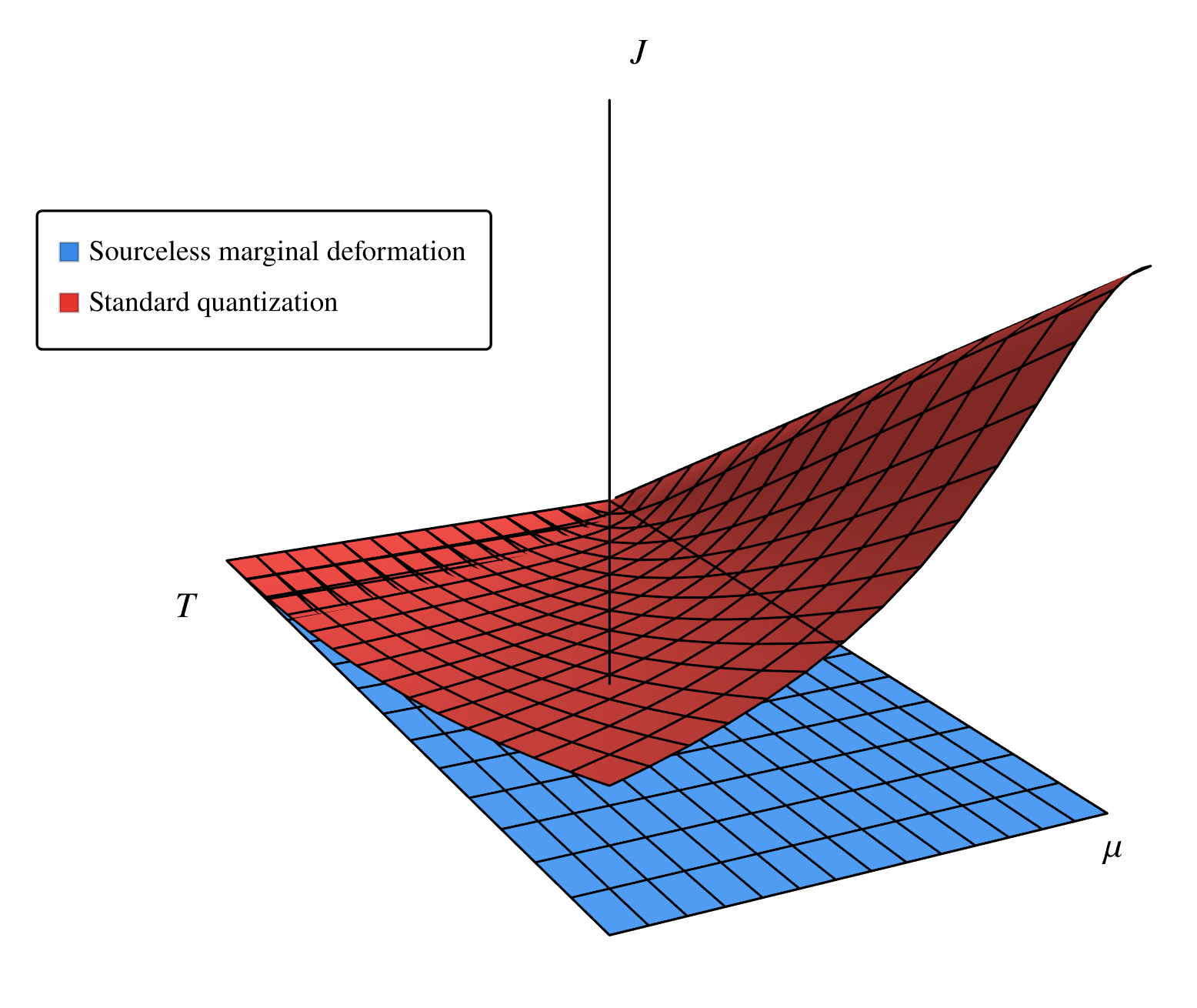}
     \caption{\ac{agr} solution manifold in the thermodynamic parameter space of source $\{T, \mu, J \}$ for two specific choices of boundary interpretations (cases  \ref{list:marginal} and \ref{list:standard}). The sourceless marginal case has trivial source and is by itself a 2-charge submanifold while the standard quantization case has a constrained source which leads to the non-trivial corrections in $s_1,\rho_1$.}
     \label{fig:plotParameterSpace}
 \end{figure}

There is, however, the special case \ref{list:marginal}. When the deformation is purely marginal and sourceless -- $a = 0$ and $b = \frac{1}{\sqrt{3}}$ -- we can immediately infer that the variations of $J = 0$ will be trivial. In that case, we will have $s_1 = s_2$ and $\rho_1 = \rho_2$. The way to understand this is that within the 3-parameter space of possible solutions quantified by $(T,\mu, J)$ the 2-parameter \ac{agr} solution spans a different subspace depending on the quantization choice for the dual boundary theory.
\autoref{fig:plotParameterSpace}, 
illustrates
how 
this difference of boundary interpretation between the alternate quantization with sourceless marginal deformation of case \ref{list:marginal} and the standard quantization of case \ref{list:standard} changes the shape of the \ac{agr} solution manifold inside the thermodynamic space of sources $\{T, \mu, J \}$. This visualization allows us to see at a glance how the sourceless marginal deformation 
reduces to a 2-charge thermodynamic space 
where 2-parameters of the solution naturally coincide with $T,\mu$ 
while the standard quantization interpretation of the \ac{agr} solution induces some non-trivial projection when varying the Gibbs free energy w.r.t. $T,\mu$.
For the sourceless marginal deformation the thermodynamics of the boundary %
thus simplifies greatly and will behave in a similar fashion to the conformal fluid dual to the \ac{rn} black hole solution.

To complete the argument above
we shall construct numerical solutions to the equations of motion \eqref{eq:eomGR} in the next section that differ from the \ac{agr} solution in that they explore the third direction orthogonal to $T,\mu$ and analyse their various boundary interpretations.

\section{Deformed Gubser-Rocha black holes}
\label{sec:oneparameterfam}

\subsection{Numerically constructed solutions}
The solutions that generically differ from \eqref{eq:homSolution} 
correspond to setting different boundary conditions for the dilaton field. 
However, for each such new solution, its interpretation depends on the quantization one considers, i.e. what the on-shell value of the action including boundary terms reads.

We will solve the \ac{gr} equations of motion \eqref{eq:eomGR} numerically
using the following parametrization
\begin{equation}
    \label{eq:parametrizationNumerics1}
    \phi = \frac{\sqrt{3}}{2} z \,\psi(z)~, \quad A_t(z) = \mu\, j(z) a_t(z)~,
\end{equation}
and with metric ansatz
\begin{equation}
    \dd s^2 = \dfrac{1}{z^2} \left[ -f(z) G_{tt}(z) \dd t^2 + \dfrac{\dd z^2}{f(z)} G_{zz}(z) + g(z) G(z) \biggl( \dd x^2 + \dd y^2 \biggr) \right]~,
\end{equation}
where $f(z),g(z),j(z)$ are held fixed to their expressions in the \ac{agr} solution \eqref{eq:homSolution} and $\psi,a_t,G_{tt}, G_{zz},G$ are the dynamical fields. The radial coordinate $z$ spans the range from the boundary at $z = 0$ to the outer horizon at $z = z_h$. The IR boundary conditions are chosen to have a single zero horizon corresponding to a non-extremal black hole and to impose regularity at the horizon for other fields (see e.g., \cite{Horowitz:2012ky}).\footnote{The boundary conditions from regularity imply in particular that $G_{tt}(z_h) = G_{zz}(z_h)$. This conveniently allows us to set the temperature with the parameters $Q$ and $z_h$ just like in the  \ac{agr} solution in Eq.~\eqref{eq:GCmatching}, as the temperature of this generalised model is given by $T = T_{\mathrm{GR}}\sqrt{ G_{tt}(z_h)/G_{zz}(z_h) } =  3\sqrt{1+Q z_h}/{4\pi z_h}$.} The UV boundary conditions are chosen to impose AdS asymptotics for the metric components and $A_t(0)=\mu$.
Parametrizing $\mu = \sqrt{3 Q z_h (1+Q z_h)}/z_h$ as in the \ac{agr} solution, the scalar boundary condition \eqref{eq:boundary-condition} can be rewritten in terms of the falloffs of $\psi$ as
\begin{equation}
    \label{eq:boundary-condition-psi}
    \psi^\prime(0) = - \dfrac{2 J}{\sqrt{3}} + \left( a - \dfrac{3 Q}{4} \right) \psi(0) + \dfrac{\sqrt{3} b}{4} \psi(0)^2~.
\end{equation}
For simplicity, we will choose $z_h = 1$ and the temperature of the solutions will therefore be encoded by $Q = \frac{3 \mu^2}{16 \pi^2 T^2}$. 
In holography, we would usually first fix the boundary theory of interest by choosing $a,b$. Then every solution to the equations of motion would be labeled by $(T, \mu, J)$ imposed through the boundary conditions. However in this section, we will be interested in how a given set of solutions, labeled by $(T, \mu, \psi(0))$, behaves in the various compatible boundary theories. This is possible because the boundary condition we impose on the scalar is simply a way to parametrize how we choose a bulk solution constrained to have a black hole in the interior. Every boundary theory determined by $a,b$ and the value of sourcing $J$ compatible with the condition \eqref{eq:boundary-condition-psi} will provide a valid boundary description. We will focus on the boundary interpretations in the next subsection. 
In many holographic studies $\psi(0)$ is often used interchangeably with the source $J$, but this is of course only true in standard quantization. We shall, however, be careful to distinguish between the boundary value $\psi(0)$ of the AdS scalar field and the source $J$ of the operator in the quantization choice dependent dual field theory.

\begin{figure}
 \centering
 \includegraphics[width=\textwidth]{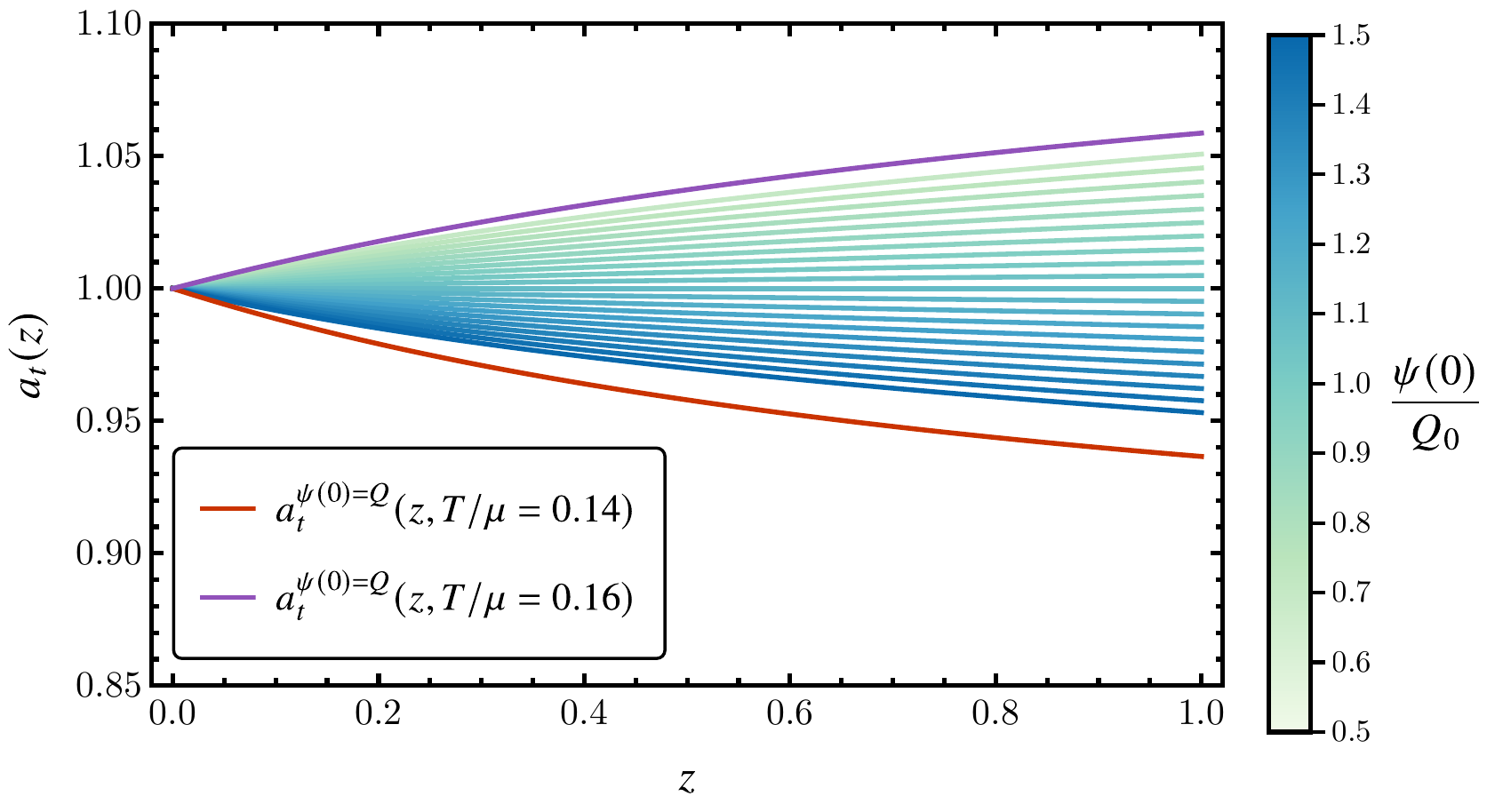}
 \caption{Gauge field component $a_t(z)$ as defined in \eqref{eq:parametrizationNumerics1} at $T_0/\mu = 0.15$ and for various values of $\psi(0)$. We compare with the equivalent function $a_t^{\psi(0) = Q}$ of the \ac{agr} solution at different temperatures $T/\mu = 0.16$ (purple) and $T/\mu = 0.14$ (red). This illustrates that qualitatively the effect of changing the dilaton boundary value has similarities to changing the ratio $T/\mu$.
 }
 \label{fig:gaugeFieldSource}
\end{figure}

Let us now briefly describe the effect of changing $\psi(0)$ without referring to any specific boundary theory. By looking at the \ac{agr} solution \eqref{eq:homSolution}, we see that $\psi(0) = Q \sim (T/\mu)^{-2}$ for this family. Therefore, increasing $\psi(0)$ is akin to lowering the temperature and vice versa. 
To confirm our intuition, we can compare solutions at fixed $Q_0 \sim (T_0/\mu)^{-2}$, and varying $\psi(0)$, to \ac{agr} solutions with $\psi(0)= Q \neq Q_0$ i.e., at different $T/\mu \neq T_0/\mu$. We will choose to focus on the gauge field $A_t(z)$ and more specifically the component $a_t(z)$ defined in \eqref{eq:parametrizationNumerics1}. 
Formally, $a_t(z) = A_t(z)/(\mu j(z, T_0/\mu))$ for a fixed $T_0/\mu$. Since the \ac{agr} solution at a different temperature $T/\mu$ will have a gauge field $A_t(z) = \mu j(z, T/\mu)$, the correct field to compare with will be $a^{\psi(0) = Q}_t(z, T/\mu\neq T_0/\mu) = j(z,  T/\mu)/j(z, T_0/\mu)$. 
We plot the profiles $a_t^{\psi(0)\neq Q_0}(z, T_0/\mu)$ in \autoref{fig:gaugeFieldSource} and compare these to $a_t^{\psi(0) = Q}(z, T/\mu > T_0/\mu)$ (purple) and $a_t^{\psi(0) = Q}(z, T/\mu<T_0/\mu)$ (red). We see that indeed, starting from $\psi(0) = Q_0$, as we increase (decrease) $\psi(0)$ with $Q_0$ fixed, the solution becomes similar to the \ac{agr} solution at lower (higher) $T/\mu$. 

\subsection{The holographic dual of the one-parameter family of solutions in different quantization choices}

Having numerically constructed instances of this 
one-parameter deformation of fixed $T/\mu$  \ac{gr} black holes, {\em each} instance in turn has multiple holographic dual interpretations depending on the quantization scheme. These are constrained by the compatibility condition \eqref{eq:boundary-condition-psi}. We will focus on three specific choices:
\begin{enumerate}
    \item \label{cases:marginal} the conformal symmetry preserving quantization $a,J = 0$ boundary theory for which we can then label our solutions by $b(\psi(0)) = \frac{4}{\sqrt{3} \psi(0)^2} \left(\psi^\prime(0) + \frac{3 Q}{4} \psi(0) \right)$,
    \item \label{cases:standard} the standard quantization boundary theory with the label $J = \alpha = \frac{3}{2} \psi(0)$,
    \item \label{cases:alternate} the alternate quantization boundary theory with $a,b = 0$ for which the label is now $J = -\beta = - \frac{3}{2} \psi^\prime(0) - \frac{3 \sqrt{3} Q}{8} \psi(0)$.
\end{enumerate}

\begin{figure}
     \centering
     \includegraphics[width=\textwidth]{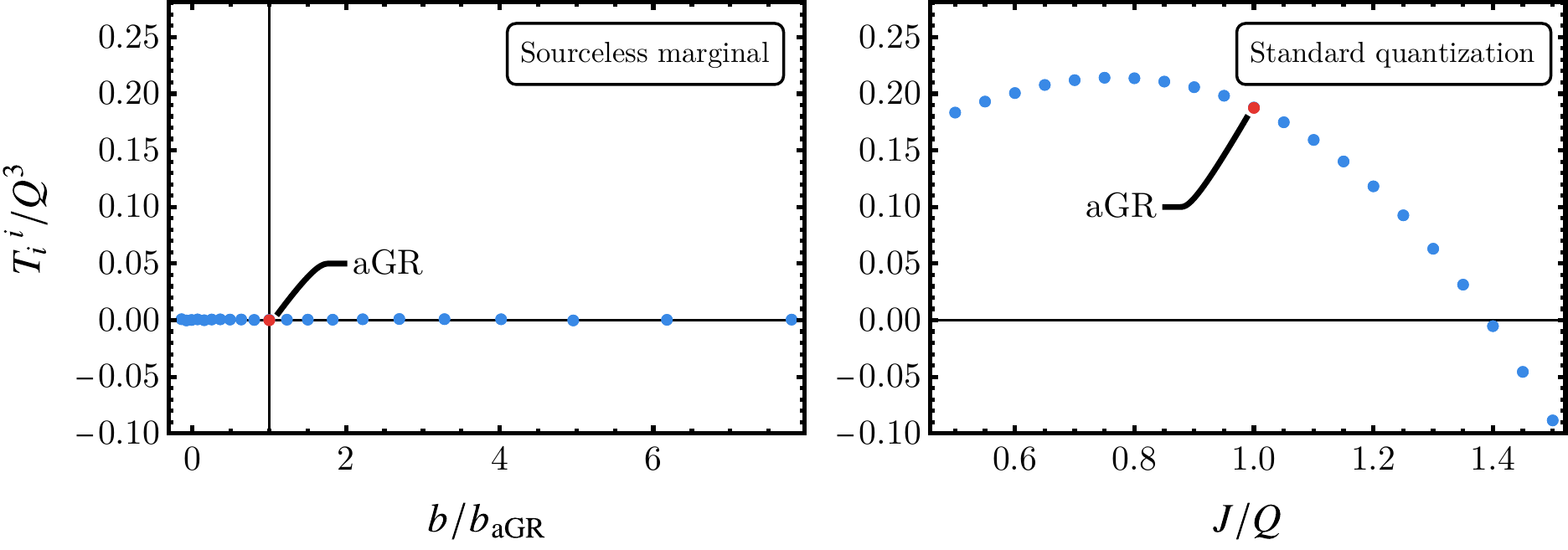}
     \caption{Trace of the boundary stress tensor when varying the Dirichlet boundary condition $\psi(0)$. This can be interpreted as exploring boundaries with $a = 0, J = 0$ and varying marginal coupling $b$ (left) or as changing the source $J = \alpha$ in standard quantization (right). aGR denotes the analytically known \acl{gr} solution. (Left) We see that in this case, $T_{ij}$ remains traceless regardless of $b$ which is consistent with a marginal deformation and the result \eqref{eq:traceSTgeneral}. (Right) In standard quantization, the trace is generically not zero, but this can happen for specific boundary theories: sourceless $J = 0$ -- not visible on the graph -- and when $\mathcal O_\varphi = 0$ -- which happens at $J/Q \simeq 1.4$).
     }
     \label{fig:traceST}
 \end{figure}

Using Eq.~\eqref{eq:vevs} we can compute the energy and the pressure of a solution in a specific quantization scheme and construct the trace of the stress tensor $T_i{}^i=-\epsilon +2P$ for each of these solutions. For the choice \ref{cases:marginal}, as we can see in \autoref{fig:traceST}, the stress tensor remains traceless for any value of $b(\psi(0))$, confirming the analytic result Eq.~\eqref{eq:traceSTgeneral}. This is what we expect from a CFT deformed by a marginal operator. On the other hand, for the choice \ref{cases:standard}, we see that generically conformality is broken and the stress tensor acquires a non zero trace. In this quantization scheme, this is also true for the \ac{agr} solution, as we described in the case \ref{list:standard}. There are two exceptions: the first one is when $J = 0$ (but $\mathcal O_\varphi \neq 0$) -- which is reminiscent of a $\mathbb Z_2$ spontaneously symmetry breaking solution but here, the finite charge of the black hole actually always leads to an explicitly symmetry broken (ESB) solution $\phi(z) \neq 0$. This case is outside the range of the plot \autoref{fig:traceST}.
The second solution would happen around $J/Q \approx 1.4$ such that $\mathcal O_\varphi = 0$. These are consistent with what we would have expected from $T_i{}^i = \alpha \beta$.

\begin{figure}
 \centering
 \includegraphics[width=\textwidth]{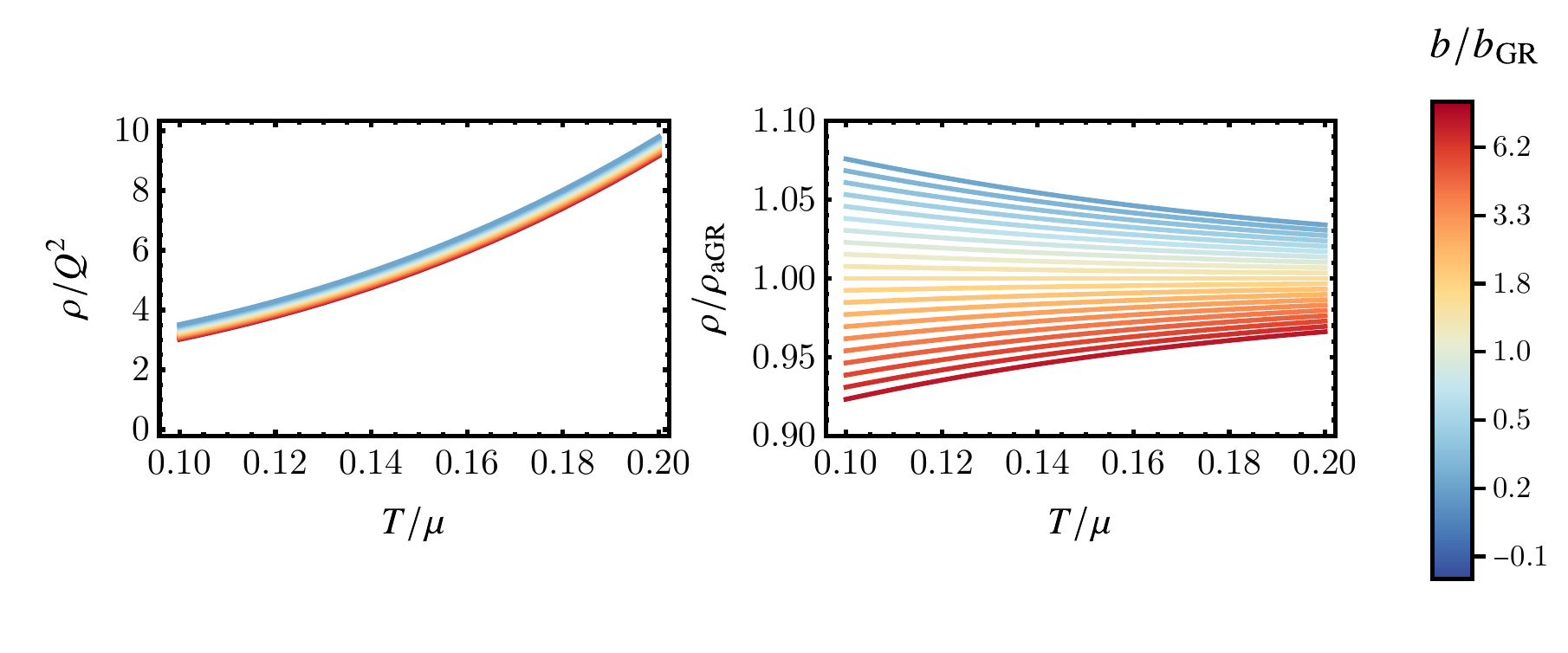}
 \caption{{Boundary charge density as a function of the temperature $T/\mu$  when imposing Dirichlet boundary conditions, which we interpret as varying the boundary theory through $b$. The charge density is normalized by $Q^2$ in the left-hand plot and by its \ac{agr} value defined in \eqref{eq:BlackHoleEntropyChargeDensity} in the right-hand plot. The qualitative behaviour of all these theories is extremely similar to the \ac{agr} solution (left) but quantitatively differs as a function of $T/\mu$ (right), showing the theories described are different.}}
 \label{fig:chargeDensityFunT}
\end{figure}

Each one of these new black hole solutions has a different thermodynamics compared to the \ac{agr} solution.
A clean way to exhibit this is to show the boundary charge density $\rho_2$, which for the choice \ref{cases:marginal} is the same as the variation of the Gibbs free energy w.r.t. the chemical potential, i.e. in that case $\rho_2=\rho_1$. 
In \autoref{fig:chargeDensityFunT}, we plot the charge density as a function of temperature for various values of the marginal coupling $b$. It is clear from this figure that the charge density as a function of $T/\mu$ is dependent on the choice of boundary theory and the deformed solution describes a different state, even if the change is small.

\begin{figure}
 \centering
 \includegraphics[width=\textwidth]{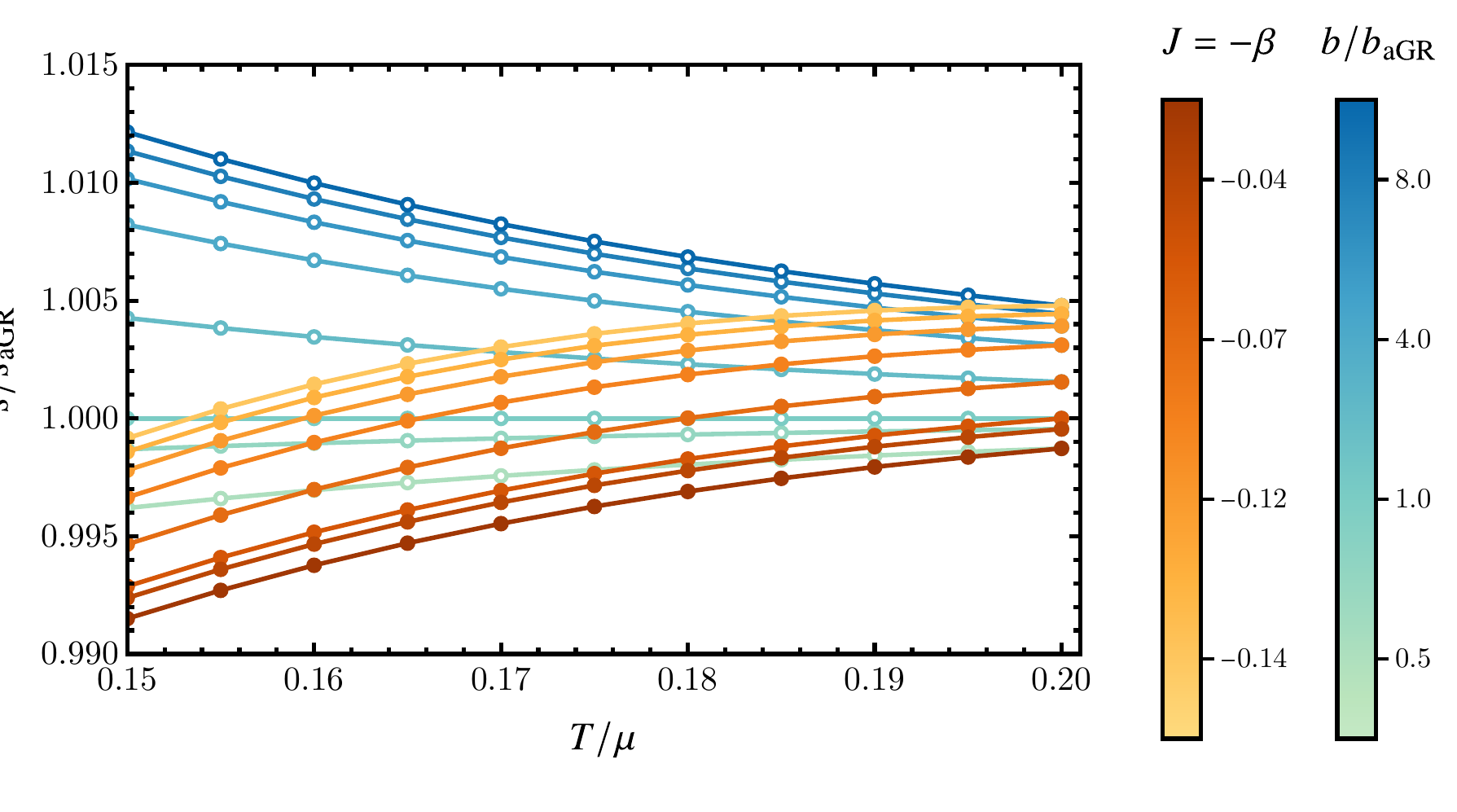}
 \caption{Black hole entropy as a function of $T/\mu$ when keeping either the alternate quantization source $J = -\beta$ fixed (choice \ref{cases:alternate}, orange gradient curves) or when keeping the label $b(\psi(0))$ fixed (choice \ref{cases:marginal}, blue gradient curves). The curves meet in pairs at $T/\mu$ -- indicating identical bulk solutions -- and separate for other temperatures -- indicating different black hole solutions.}
 \label{fig:entropyVariousTheories}
\end{figure}

To reiterate this last point, let us remember that a priori, the true charge density of the theory $\rho_2$, as well as the true entropy of the theory $s_2$, only depend on the bulk solution -- they are geometric quantities. Yet we now argue that different boundary theories have different thermodynamics. The resolution of this apparent contradiction is that while the entropy and charge density of a black hole solution only really depend on the bulk solution, how we explore the space of solutions is dependent on the choice of quantization. As we mentioned in Section~\ref{sec:oneparameterfam}, 
the holographic interpretation of black hole thermodynamics shows that we should
label solutions by their sources $\{T, \mu, J\}$ -- and in the case of the sourceless solutions of the choice \ref{cases:marginal}, $b$ plays the role of the label $J$. But different boundary theories have different notion of source $J$ such that varying $T$ and $\mu$ at fixed $J$ will mean different path in the space of bulk solutions labeled by $\{T, \mu, \psi(0)\}$. In \autoref{fig:entropyVariousTheories}, we illustrate this point by looking at the Bekenstein-Hawking entropy $s_2$ as a function of $T/\mu$ -- all solutions are normalized by the \ac{agr} entropy defined in \eqref{eq:BlackHoleEntropyChargeDensity}. Both choices \ref{cases:marginal} and \ref{cases:alternate} are used to label the solutions when varying the temperature, which can be done by imposing the boundary condition \eqref{eq:boundary-condition-psi} for each of the choices. The values of $b(\psi(0))$ and $J = -\beta$ are chosen such that solutions meet in pair at $T/\mu = 0.2$. Upon lowering the temperature, we see that these pairs split indicating that the bulk solutions they belong to are not the same anymore. A path at fixed $J = -\beta$ is therefore generically different than a path at fixed $J = \alpha$ or fixed $b(\psi(0))$.

\FloatBarrier
\section{Conclusion}

In this paper, we have clarified how the \ac{gr} black hole thermodynamics works in the context of holography and the appropriate quantization thereof. The well-known analytical solution~\eqref{eq:homSolution} of~\cite{Gubser:2009qt} covers only a 2-parameter subspace of the full 3-parameter thermodynamics of black hole solutions to the action~\eqref{eq:gravAction}. The 2-parameter \ac{agr} black hole solution has been used widely as a physically sound version of the $z=\infty$ AdS$_2$ IR critical point that preserves the quantum critical properties but does so with a vanishing zero temperature entropy. It was already pointed out \cite{Kim:2016dik} that an unusual quantization choice could preserve conformal thermodynamics and hence stay within the analytically known 2-parameter family. This indicates the existence of a marginal operator in this specific quantization scheme \cite{Caldarelli:2016nni} and we have recovered this in our analysis. For other quantization choices, the analytic solution has a fine tuned value for the source. To prove this point we have numerically computed the solutions corresponding to different boundary values of the dilaton. This fills out the full 3-parameter thermodynamic phase space.
The filled out phase-space therefore elucidates that other quantization choices are just as valid as the one we chose to focus on. This had to be so, but the trade-off that one must make is to properly account for various scalar contributions to the general thermodynamics of the theory in line with the findings in \cite{Li:2020spf}. 

Because the \ac{gr} action is a consistent truncation of $d=11$ supergravity compactified on AdS$_4\times S_7$ and has ABJM theory as its known holographically dual CFT, in principle one should be able to identify this marginal operator in the CFT. The fact that marginality is associated with a multitrace deformation makes this not as straightforward as may seem. In particular as it originates naturally in alternate quantization, it is likely that it is an operator which is only marginal in the large $N$ limit where the classical gravity description applies. We leave this for future research.

Our focus and interest is the use of the \ac{gr} and other \ac{emd} models as phenomenological descriptions of AdS$_2$ fixed points, especially due to its resemblance to the experimental phenomenology of strange metals. In this comparison, thermodynamic susceptibilities and (hydrodynamic) transport play an important role. Our result here shows that in \ac{emd} models one must be precise in the choice of boundary conditions and scalar quantization as they will directly affect the long-wavelength regime of the dual boundary theory as well as correct the thermodynamics of any extension of the \ac{gr} model. This is especially true for any boundary interpretation differing from the pure marginal case of \cite{Kim:2016dik,Caldarelli:2016nni}, as was shown by \cite{Li:2020spf} for Einstein-Scalar models and we have shown here for the \ac{gr} model. A proper understanding of the boundary conditions is necessary both for the thermodynamics of the background and the hydrodynamic fluctuations on top of that background.

\section*{Acknowledgements}

We thank J. Aretz, R. Davison, K. Grosvenor,  J. Zaanen and especially A. Krikun for
discussions during the Nordita scientific program {\em Recent Developments in Strongly Correlated Quantum Matter}. This research was supported in part by the Dutch Research Council (NWO) project 680-91-116 ({\em Planckian Dissipation and Quantum Thermalisation: From Black Hole Answers to Strange Metal Questions.}), the FOM/NWO program 167 ({\em Strange Metals}), and by the Dutch Research Council/Ministry of Education. 

\appendix
\section{Validity of the boundary action}
\label{sec:discontinuous-field-redefinition}

In a previous version of this paper, we considered the boundary term introduced by \cite{Kim:2016dik} which is of the form
\begin{equation}
    \label{eq:bdyPhiKim}
    S^{(c_\phi)}_{\bdy,\phi} = \int_{z = \epsilon} \dd^3 x \, \sqrt{-\gamma} \left[ \frac{\Lambda_\phi}{2} \phi^2 + c_\phi \phi N^z \partial_z \phi \right]~, \quad \Lambda_\phi = 2 c_\phi - 1~,
\end{equation}
which matches our boundary terms for specific values $S_{\bdy, \phi}^{(c_\phi = 0)} = S_{\bdy, \phi}^{(\mathrm{SQ})}$ and $S_{\bdy, \phi}^{(c_\phi = 1)} = S_{\bdy, \phi}^{(\mathrm{MT})}$ for $a = 0, b = 0$. The claim of \cite{Kim:2016dik} is that more general values of $c_\phi$ are also possible, which from a renormalization point of view is an acceptable assumption. The only prescription one has for boundary terms is to choose relevant and marginal ones (the irrelevant boundary terms contribute as corrections in the cutoff $\epsilon$ and can be truncated) which respect the symmetries of the action. However, choosing the boundary term \eqref{eq:bdyPhiKim} leads to
\begin{equation}
    \label{eq:variationKim}
    \begin{aligned}
        \delta \left( S_{\bulk} + S^{(c_\phi)}_{\bdy,\phi} \right) & = \int_{z = \epsilon} \dd^3 x \sqrt{-\gamma} \, \left[ (1 - c_\phi) \beta \delta \alpha - c_\phi \alpha \delta \beta \right]\\
        & = \int_{z = \epsilon} \dd^3 x \sqrt{-\gamma} \, \left(- c_\phi \alpha^{1/c_\phi} \right) \delta \left( \beta \alpha^{1 - 1/c_\phi} \right)
    \end{aligned}
\end{equation}
which generically differs from our result for the standard quantization or multi-trace deformation where $\mathcal O_\varphi = \alpha \text{ or } \beta$.

The question of the validity of such variational problem as Eq.~\eqref{eq:variationKim} was raised before in e.g. \cite{Dyer:2008hb} for the simple case of a non-relativistic particle. 
Consider a particle with action $S_1 = \int_{t_1}^{t_2} \dd t (-\dot q^2/2)$ to which one adds the total derivative term $S_2 = \left[\frac{1}{2} q \dot q\right]_{t_1}^{t_2}$. The variation of the total action on-shell $\delta(S_1 + S_2) = \left[ \frac{1}{2} q \delta \dot q - \frac{1}{2} \dot q \delta q \right]_{t_1}^{t_2}$ is of a similar form as the variation \eqref{eq:variationKim} for $c_\phi = 1/2$. The boundary condition required to make the boundary variation well-defined is then to fix $\dot q/q = C$ at $t = t_1$ and $t = t_2$. However, in the case of $S_1$, this is not a correct boundary condition to impose. Since the bulk equation of motion is $\ddot q = 0$ with solutions $q(t) = A t + B$ and $\dot q(t) = A$, the quantity to fix is $\frac{\dot q}{q} = \frac{A}{A t + B} = \frac{1}{t + B/A}$ which only depends on the ratio $B/A$. Therefore, fixing it at $t_1$ leaves no freedom to also fix it at $t_2$. At the same time the two boundary conditions at $t_1$ and $t_2$ do not select a unique solution. A direct check one can do is whether for other values of the analogous $c_\phi$, this problem remains. Taking for example $S_2 = \left[\frac{1}{3} q \dot q\right]_{t_1}^{t_2}$, the boundary condition to impose is now to fix $\dot q/q^2 = \frac{A}{(A t + B)^2}$. Solving this condition at the boundaries for values $C_{1,2}$ now does lead to fully determined solutions, unlike the previous case. However, the solutions are not unique, because the boundary conditions itself have  arbitrary constants $C_{1,2}$. There are therefore 
multiple branches to the system of equations $A C_{1,2} = (A t_{1,2} + B)^2$.

In holography only the UV boundary conditions are imposed in the exact same manner. The IR boundary condition in a black hole spacetime is different. We simply require regularity of the scalar at the event horizon. 
For $c_\phi = 1/n$,  $n\in \mathbb N^*$, the question of whether the variational problem is well-defined is then whether the UV boundary condition of fixing $\frac{\beta}{\alpha^{n-1}} = C$ is sufficient to pick a unique solution once the IR boundary conditions are taken into account. 
It is quite straightforward to show that these are the same boundary conditions as the usual multi-trace deformation boundary condition \eqref{eq:compatibility-general}, for $J = 0$ and specific choices of monomial $\mathcal F_n = \frac{a_n}{n} \alpha^n$. From \eqref{eq:boundary-condition}, we see that for $n > 1$, the sourceless boundary condition for the deformation associated with $\mathcal F_n$ is $\frac{\beta}{\alpha^{n-1}} = \frac{a_n}{n-1}$ so the matching between boundary theories occurs for $C = \frac{a_n}{n-1}$. Interestingly, choosing the boundary value $C$ is equivalent to choosing a deformation coupling constant with (single-trace) scalar source $J = 0$. This is because the coupling constant $a_n$ is really the same as a source for the multi-trace operator $\mathcal O^n$.

In Table~\ref{tbl:analog-bdy-conditions} we look 
at 
$n=1,2,3,\infty$ and what type of multi-trace deformation they match. For $n \geq 4$
the higher order terms in $\mathcal F$ represent 
irrelevant operators and we shall not consider them.
The special 
cases $n = 1$ and $n = \infty$ i.e. $c_\phi = 1$ and $c_\phi = 0$ 
are the alternate and standard quantization case of fixing $\alpha = J$ and $\beta = -J$. In the previous version of this article we argued that the \ac{agr} solution quantized with boundary term \eqref{eq:bdyPhiKim} and $c_\phi=1/3$ could be viewed as a marginal deformation with $n = 3$ and $\beta/\alpha^2 = \frac{1}{2 \sqrt{3}}$ which according to our mapping is equivalent to the case \ref{list:marginal}, as expected.

\begin{table}[h]
\begin{center}
\begin{tabular}{ |c|c|l l l| }
    \hline
    $n$ & Boundary condition & \multicolumn{3}{c|}{Analog multi-trace choice}\\
    \hline
    $n = 1$      & $\beta = C$ & $a = 0$, & $b = 0$, & $J = -C$\\[1.5pt]
    $n = 2$      & $\frac{\beta}{\alpha} = C$ & $a = C$, & $b = 0$, & $J = 0$\\[1.5pt]
    $n = 3$      & $\frac{\beta}{\alpha^2} = C$ & $a = 0$, & $b = 2C$, & $J = 0$\\[1.5pt]
    $n = \infty$      & $\alpha = C$ & $a = 0$, & $b = 0$, & $J = C$\\[1.5pt]
    \hline
\end{tabular}
\end{center}
\caption{Matching between the boundary conditions obtained from the multi-trace deformation boundary action \eqref{eq:bdyPhiMT} and those obtained from the boundary term \eqref{eq:bdyPhiKim}.}
\label{tbl:analog-bdy-conditions}
\end{table}

Moreover, and importantly, the on-shell values of the boundary actions \eqref{eq:bdyPhiMT} with monomial multitrace deformations $\mathcal F = \mathcal F_n$ and \eqref{eq:bdyPhiKim} are also equivalent through the mapping described in Table~\ref{tbl:analog-bdy-conditions}. Indeed, we see that the difference between the boundary terms is
\begin{equation}
    \label{eq:differenceMTKim}
    S_{\bdy,\phi}^{(\rm{MT})}(\mathcal F = \mathcal F_n) - S_{\bdy,\phi}^{(c_\phi)} = \int_{z = \epsilon} \left[ \dfrac{a_n}{n}\alpha^n - (1-c_\phi) \alpha \beta \right] = \int_{z = \epsilon} \left[ a_n - C (n-1) \right] \frac{\alpha^n}{n}~,
\end{equation}
where we injected the expansion $\phi \sim \alpha z + \beta z^2$ and in the second equality, we used the boundary condition $\beta = C \alpha^{n-1}$ with $c_\phi = 1/n$. We see that the difference \eqref{eq:differenceMTKim} vanishes for the choice $C = \frac{a_n}{n-1}$ and thus the actions are the same through the mapping described in Table~\ref{tbl:analog-bdy-conditions}. We can conclude that as far the two roles of the boundary terms go -- setting the boundary conditions of the variational problem and specifying an on-shell value for the action -- these boundary terms yield the same answer for specific choices of the boundary theory. This explains how our previous derivation based on \eqref{eq:bdyPhiKim} yielded the same results as the derivation based on \eqref{eq:bdyPhiMT} for sourceless solutions. The on-shell action equivalence does not hold in generality, however. The boundary term \eqref{eq:bdyPhiKim} fails to account for polynomial deformations $\mathcal F$ and therefore would miss out on the most general theories of case \ref{list:generic}. 

\section{Matching of metric gauge choices}
\label{sec:match-fg}

In Eq.~\eqref{eq:scalarExp} we have expressed our scalar field UV expansion in the \ac{fg} gauge choice for the metric \eqref{eq:fg-ansatz}. In this section we will use $r$ to denote this choice of radial coordinate. However, the \ac{agr} solution \eqref{eq:homSolution} uses a different metric gauge choice \eqref{eq:metric}. This means that the expansion of the scalar field $\phi = \hat \alpha z + \hat \beta z^2 + \ldots$ in the \eqref{eq:metric} coordinates is not directly identical to that given in Eq.~\eqref{eq:scalarExp}. 
They are related by
solving $\frac{\dd r^2}{r^2} = \frac{\dd z^2}{z^2 f(z)}$. 
This relation is formally given by
\begin{equation}
    \label{eq:relation-radial-coordinates}
    \log r(z) - \log \epsilon = \int_\epsilon^z \dfrac{\dd x}{x \sqrt{f(x)}}~, \quad \text{with } \epsilon \to 0~.
\end{equation}
In the near-boundary regime, we will only be interested in the leading and subleading orders of this relation -- since we only want to see how the leading and subleading orders in the scalar expansion mix -- and we therefore expand $f(z) = 1 + f^\prime(0) z + \ldots$, where the analytical value of $f$ is given in Eq.~\eqref{eq:homSolution}. Doing so, we find
\begin{equation}
    \label{eq:relation-radial-coordinates-exp}
    r(z) \sim z - \dfrac{3 Q z^2}{4} + \mathcal{O}(z^3)~.
\end{equation}
It is then straightforward to input this in the \ac{fg} UV expansion
\begin{equation}
    \phi \sim \alpha r + \beta r^2 \sim \alpha z + \left( \beta - \dfrac{3 Q}{4} \alpha \right) z^2~,
\end{equation}
as was claimed in Eq.~\eqref{eq:scalarToFG}.

\bibliographystyle{unsrt}
\bibliography{refs}

\end{document}